\documentclass[submission,copyright,creativecommons]{eptcs}
\providecommand{\event}{MOD$^*$ 2014}

\usepackage{latexsym,amssymb,url,boxedminipage}
\usepackage{amsmath}
\usepackage{amssymb}
 \usepackage{times}
 \usepackage{enumerate}

\newcommand{\transI}{transI}
\newcommand{\transF}{transF}

\newcommand{\chor}[3]{#1_{#2\rightarrow #3}}

\newcommand{\pinull}{\mathbf{0}}

\newcommand{\parop}{\,|\,}      

\makeatletter
\def \rightarrowfill{\m@th\mathord{\smash-}\mkern-6mu%
  \cleaders\hbox{$\mkern-2mu\mathord{\smash-}\mkern-2mu$}\hfill
  \mkern-6mu\mathord\rightarrow}
\makeatother

\newcommand{\bigfract}[2]{\frac{^{\textstyle #1}}{_{\textstyle #2}}}

\newcommand{\ntrans}[1]{\stackrel{#1}{\longrightarrow}\!\!\!\!\!\!\!\! /\ \  }
\newcommand{\trans}[1]{\stackrel{#1}{\longrightarrow}}

\newcommand{\pisucc}{\mathbf{1}}
\newcommand{\re}{{\setminus \!\!\!\!\; \setminus}}

\newcommand{\pa}{{|\!|}}
\newcommand{\vuota}{\varepsilon}
\newcommand{\sem}[1]{[\![#1]\!]}

\newtheorem{adefinizione}{Definition}[section]
\newtheorem{fatto}[adefinizione]{Fact}
\newtheorem{alemma}[adefinizione]{Lemma}
\newtheorem{teorema}[adefinizione]{Theorem}
\newtheorem{corollario}[adefinizione]{Corollary}
\newtheorem{proposizione}[adefinizione]{Proposition}
\newtheorem{esempio}[adefinizione]{Example}

\newenvironment{definition}{\begin{adefinizione}\ \rm}{\end{adefinizione}}

\newenvironment{theorem}{\begin{teorema}\ \rm}{\end{teorema}}

\newenvironment{proposition}{\begin{proposizione}\ \rm}{\end{proposizione}}
\newenvironment{example}{\begin{esempio}\ \rm}{\end{esempio}}

\title{Choreographies and Behavioural Contracts on the Way to Dynamic Updates}
\author{Mario Bravetti \ \ \ Gianluigi Zavattaro
\institute{University of Bologna, Italy / INRIA, France}
\email{\{mario.bravetti,gianluigi.zavattaro\}@unibo.it}
}

\def\titlerunning{Choreographies and Behavioural Contracts on the Way to Dynamic Updates}
\def\authorrunning{M. Bravetti and G. Zavattaro}

\begin{document}
\maketitle
\begin{abstract}
We survey our work on choreographies and behavioural contracts in multiparty interactions.
In particular theories of behavioural contracts are presented which enable reasoning about correct service
composition (contract compliance) and service substitutability (contract refinement preorder) under different
assumptions concerning service communication: synchronous address or name based communication with
patient non-preemptable or impatient invocations, or asynchronous communication.
Correspondingly relations between behavioural contracts and choreographic descriptions are considered,
where a contract for each communicating party is, e.g., derived by projection.
The considered relations are induced as the maximal preoders which preserve contract compliance and global traces:
we show maximality to hold (permitting services to be discovered/substituted independently for each party) when
contract refinement preorders with all the above asymmetric communication means are considered and, instead, not to
hold if the standard symmetric CCS/pi-calculus communication is considered (or when directly relating
choreographies to behavioral contracts via a preorder, no matter the communication mean). The obtained maximal
preorders are then characterized in terms of a new form of testing, called compliance testing, where not only tests must succeed but also the
system under test (thus relating to controllability theory), and compared with classical preorders such as may/must
testing, trace inclusion, etc.
Finally, recent work about adaptable choreographies and behavioural contracts is presented, where the
theory above is extended to update mechanisms allowing choreographies/contracts to be modified at run-time
by internal (self-adaptation) or external intervention.
%
\end{abstract}

\section{Introduction}

We survey our theoretical studies about coordination and design of service oriented
systems based on the notion of behavioural contract, which describes the
interactive behaviour of a service as, e.g., obtained via session/behavioural types [8] or expressed in
orchestration languages like WS-BPEL~\cite{wsbpel-spec}.

Two main approaches are commonly considered for the composition of services: 
service {\em orchestration} and service {\em choreography}.
In an orchestration language, e.g. WS-BPEL~\cite{wsbpel-spec}, the activities of
the composed services are expressed from the viewpoint of a
specific component, called the orchestrator,
that is responsible for invoking the composed
services and collect their responses.
Choreographies languages, e.g. WS-CDL~\cite{wscdl-spec}, instead, support a high level description of peer-to-peer
interactions among several services playing different roles (multi-party interaction), from a top-level abstract viewpoint.

When implementing an interaction-oriented 
choreography by assembling
already available services, several mechanisms and notions need to be introduced.
Often the possibility is considered of extracting, from the
global specification, the orchestrational behaviour of each of the communicating parties in the form of a {\it behavioural contract} or of an abstract workflow, as expressed e.g. in abstract 
WS-BPEL~\cite{wsbpel-spec}. 
One of the first articles that relates
Choreography to Orchestrational descriptions is~\cite{sac05}. It introduces ideas concerning several technical aspects of
such a relation using e-commerce as a case study.
The extraction of choreographical behaviours is often called ``projection of the choreography on the roles'', see, e.g.~\cite{SC07}. 
The idea is that, based on the contracts derived with such a projection, services are retrieved that expose a behaviour which
is compatible with the extracted behaviours.

It is worth noticing that despite choreography projection is a powerful tool 
for relating global choreographic descriptions with local peer behaviours, 
it is not always guaranteed to be applicable. 
For instance, suppose a choreography requires an interaction to occur before
another one. If the two interactions involve distinct roles (that do 
not have other synchronisations) they could occur in any order.
For this reason, choreography projections are usually applied 
to the so-called {\em connected} choreographies, where such misbehaviours
cannot be expressed.


Another difficulty encountered when 
choreography languages 
are used to describe the message exchange among
services, is that it is not trivial to relate the high level
choreography description with the actual implementation
of the specified system realized as composition of services (orchestrations): 
they are usually loosely coupled, independently developed
by different companies, and autonomous.
More precisely, the difficult task is, given a choreography, 
to lookup available services that, once combined, are ensured to 
behave according to the given choreography.

To this aim a foundational study about coarsest refinement preorders
between behavioural contracts made it possible to decide whether a service discovered on the internet can be used to
play the role of a service with a given desired contract in the context of a multi-party coordination
(independently of the service discovered for the other roles) 
\cite{sensoria,mscs09,sfm09,fundaInfo08,coord07,fsen07}, or to play a certain role in a
given choreography \cite{libroCubo,wsfm08,SC07}, without incurring in deadlocks (and furthermore guaranteeing termination under
fairness assumption). More precisely,
theories of behavioural contracts have been introduced which enable reasoning about correct service
composition (contract compliance) and service substitutability (contract refinement preorder) under different
assumptions concerning service communication: synchronous 
address \cite{libroCubo,sensoria,SC07} or name based \cite{sfm09,fundaInfo08,fsen07} communication (invocations are directed to a certain role or just based on channel names as in CCS) with patient non-preemptable or impatient invocations (waiting admitted when invoking services, as in CCS communication, or not admitted, as for the ready-send primitive of the Message Passing Interface) \cite{mscs09,coord07}, or asynchronous communication with queues \cite{libroCubo,wsfm08}.

Concerning contract compliance, we consider a service (contract) composition to be correct if any possible execution leads to 
successful termination under fairness assumption, i.e. for any finite behavioural path of the system there exists a finite path from the reached state that leads all services to successful termination.
This means that there can be an infinite computation, but given
any possible prefix of this infinite computation, it must be possible to extend it 
to reach a successfully terminated computation.
This guarantees that the system is both deadlock and, under the fairness assumption (i.e. whenever a choice is traversed infinitely often every possible outcome is chosen infinitely often), live-lock free.

The considered refinement relations are, then, induced as the maximal preoders which preserve contract compliance:
we show maximality to hold (permitting services to be discovered/substituted independently for each party) when
contract refinement preorders with all the above asymmetric communication means are considered and, instead, not to
hold if the standard symmetric CCS/pi-calculus communication is considered (or when directly relating
choreographies to behavioral contracts via a preorder, no matter the communication mean). 

The obtained maximal
preorders are then characterized in terms of a new form of testing, called {\it compliance testing}, where not only tests must succeed but also the
system under test. The structure of this new form of testing directly relates contract refinement theory with controllability theory, in that an uncontrollable contract (i.e. a contract that cannot be led to success by any test) is equivalent to any other contract.

Moreover, in
the synchronous case, the obtained preorders have been classified with respect to standard preoreders, turning out to be coarser than standard (fair) testing preorder~\cite{RV05} and incomparable with respect to trace inclusion due to uncontrollable contracts \cite{fundaInfo08,fsen07}. In the synchronous case, we also provide 
sound characterizations of the maximal contract refinement preorder that are decidable, by resorting to 
an encoding into fair testing preorder~\cite{RV05}.

We also consider the problem of directly relating discovered behavioral contracts with choreographies: {\it choreography
conformance} relation. In this case we show that a maximal relation does not exist and we present a notion
of choreography conformance, called {\it consonance} base on combining choreography projection and the testing characterization.

%

Finally, we discuss a recent approach~\cite{beat} for extending the presented choreography and orchestration languages
with mechanisms for dynamic updates. The idea is to support both internal self-adaptations
and external modifications. The former mechanism is used to specify systems able to detect events that
require to modify the usual behaviour; the latter allows for system's modifications that are statically unpredictable.
In both cases choreographies are extended with a scope construct used to delimit those parts of the choreographies 
that can be modified at run-time; an additional update primitive injects new code inside scopes.
Scope projections are used to control which parts of the peers should be modified when an update
is executed at run-time.

The paper is structured as follows.
In Section 2 we introduce our formalization of choreographies and orchestrations 
and we relate them via the notions
of choreography implementation, projection and well-formedness. Moreover we present
the notion of Connected Choreographies.
In Section 3 we present contract-based service discovery and we report about the theory of behavioural 
contracts and refinement and its characterization in terms of testing. We also deal with the problem
of directly relating discovered behavioral contracts with choreographies. Such a theory is presented for the
case of synchronous address based communication (where invocations are directed to a certain role) with
patient non-preemptable invocations \cite{libroCubo,sensoria}. The results for other forms of communications are just summarized here.
In Section 4 recent work about adaptable choreographies and behavioural contracts is presented~\cite{beat}, where the
theory above is extended to update mechanisms allowing choreographies/contracts to be modified at run-time.
Finally, Section 5 reports some conclusive remarks about
comparison with related literature and future work.

\section{Choreographies and Orchestrations}

We start by presenting the choreography and orchestration calculi (representing individual service behaviour)
and then we relate them by projection.

\subsection{The Choreography Calculus}


We assume a denumerable set of action names ${\cal N}$, ranged over by  $a,b,c,\dots$ and 
a set of roles {Roles} ranged over by $r, s, l$.

\begin{definition}{\bf (Choreographies)}
The set of $Choreographies$, ranged over by $H, L, \cdots$
is defined by the following grammar:  
\[
H \ \ ::= \ \ \quad a_{r \rightarrow s} \quad | \quad
                    H+H \quad | \quad
                    H;H \quad | \quad
                    H|H \quad | \quad
                    H^*  
\]
The invocations $a_{r \rightarrow s}$ means 
that role $r$ invokes the operation $a$ provided
by the role $s$. The other operators are 
choice $\_+\_$, sequential $\_;\_$, parallel $\_|\_$, 
and repetition $\_^*$.
\end{definition}


The operational semantics of choreographies
considers two auxiliary terms
$\pisucc$ and $\pinull$. They are used to model
the completion of a choreography, which is relevant
in the operational modeling of sequential composition.
The formal definition is given in Table~\ref{choreosem} 
where we take $\eta$ to range over the set of 
labels $\{ a_{r \rightarrow s}\ |\ a \in {\cal N}, r,s \in Roles \} \cup \{ \surd\}$
(the label $\surd$ denotes successful completion). 
The rules in Table~\ref{choreosem} are rather standard 
for process calculi with sequential composition
and without synchronization; in fact,
parallel composition simply allows for the interleaving
of the actions executed by the operands (apart from completion labels $\surd$
that are required to synchronize).

\begin{table}[th]
\[
\begin{array}{ccc}
a_{r \rightarrow s} \trans{a_{r \rightarrow s}} \pisucc 
&
\pisucc \trans{\surd} \pinull 
&
H^* \trans{\surd} \pinull 
\\
\\
\bigfract{H \trans{\eta} H'}
         {H\!+\!L \trans{\eta} H'} 
&         
\bigfract{H \trans{\eta} H' \quad \eta \neq \surd} 
         {H;\!L \trans{\eta} H';\!L}                  
\qquad
&
\bigfract{H \trans{\surd} H' \quad L \trans{\eta} L'}
         {H;\!L \trans{\eta} L'}        
\\
\\
\bigfract{H \trans{\surd} H' \quad L \trans{\surd} L'}
         {H \!\parop\! L \trans{\surd} H' \!\parop\! L'}  
\qquad
&
\bigfract{H \trans{\eta} H' \quad \eta \neq \surd}
         {H \!\parop\! L \trans{\eta} H' \!\parop\! L}                                     
&
\bigfract{H \trans{\eta} H' \quad \eta \neq \surd}
         {H^* \trans{\eta} H' ; H^*}  
\end{array}
\]

\caption{Semantic rules for contracts (symmetric rules omitted).}
\label{choreosem}
\end{table}

Choreographies are especially useful to describe the protocols
of interactions within a group of collaborating services.
To clarify this point, we present a simple example of a protocol 
described with our choreography calculus.


\begin{example}\label{BSB}
{\bf (Buyer/Seller/Bank)}
Let us consider the following choreography composed of three
roles: {\em Buyer}, {\em Seller} and {\em Bank}
\[
\begin{array}{l}
Request_{Buyer \rightarrow Seller};
(\ Offer_{Seller \rightarrow Buyer} \parop
PayDescr_{Seller \rightarrow Bank}\ ); \\
\hspace{.5cm} Payment_{Buyer \rightarrow Bank};
(\ Confirm_{Bank \rightarrow Seller} \parop 
Receipt_{Bank \rightarrow Buyer}\ )
\end{array}
\]
According to this choreography, the {\em Buyer} initially
sends an item request to the {\em Seller} that subsequently,
in parallel, replies with an offer and sends a payment description to the {\em Bank}.
Then the {\em Buyer} performs the actual payment to the {\em Bank}. The latter in parallel
replies with a receipt and sends a payment confirmation to the {\em Seller}.
\end{example}

\subsection{The Orchestration Calculus}

As for choreographies, we assume a denumerable set of action names ${\cal N}$, ranged over by  $a,b,c,\dots$. We use $\tau \notin {\cal N}$ to denote an internal (unsynchronizable) computation. 

%
%

\begin{definition}\label{orcsys} {\bf (Orchestrations and Systems)}
The syntax of orchestrations is defined by the following
grammar 
\[
\begin{array}{lll}
C &\ ::=\ &      \pinull \quad | \quad 
               \pisucc \quad | \quad
               \tau  \quad | \quad
               a  \quad | \quad
		  \overline{a}_l \quad | \quad               
\\[.2cm]
  & &             
               C ; C \quad | \quad
               C\!+\!C \quad | \quad
               C \!\parop\! C \quad | \quad
               C^* 
\end{array}
\]
The set of all orchestrations $C$ is denoted by ${\cal P}_{orc}$.
\\
The syntax of systems (orchestration compositions) is defined by the following
grammar 
\[
P \ \ ::= \ \   [C]_l \quad | \quad
               	P \pa P \quad 
\]
A system $P$ is well-formed if: $(i)$ every orchestration subterm 
$[C]_l$ occurs in $P$ at a different role $l$ and $(ii)$ no output action 
with destination $l$ is syntactically included inside an orchestration subterm
occurring in $P$ at the same role $l$, i.e. actions $\overline{a}_l$ cannot occur inside a subterm $[C]_l$
of $P$.
The set of all well-formed systems $P$ is denoted by ${\cal P}$.
In the following we will just consider well-formed systems and, for simplicity, we will call them
just systems. 
\end{definition}
We take $\alpha$ to range over
the set of syntactical actions $SAct = {\cal N} \cup \{ \overline{a}_l \; | \; a \in {\cal N} \wedge l \in {Roles} \} \cup \{ \tau \}$.

The operational semantics of contracts is defined by the rules
in Table~\ref{contsem} (plus symmetric rules). The operational semantics of systems is defined by the rules
in Table~\ref{contcompsem} plus symmetric rules.
We take $\beta$ to range over
the set $Act$ of actions executable by contracts and systems, i.e.
$Act =  SAct \cup
\{ a_l \; | \; a \in {\cal N} \wedge l \in {Roles} \} \cup
\{ a_{r \rightarrow s}\ |\ a \in {\cal N} \wedge r,s \in {Roles} \} \cup 
\{ \overline{a}_{r s}\ |\ a \in {\cal N} \wedge r,s \in {Roles} \}$.
We take $\lambda$ to range over the set of 
transition labels ${\cal L} = Act \cup \{ \surd \}$, where $\surd$
denotes successful termination. 
\begin{table}[tbh]
\[
\begin{array}{cc}
\pisucc \trans{\surd} \pinull 
&
\alpha \trans{\alpha} \pisucc 
\\
\\
\multicolumn{2}{c}{
\bigfract{C \trans{\lambda} C'}
         {C\!+\!D \trans{\lambda} C'} 
\qquad         
\bigfract{C \trans{\lambda} C' \quad \lambda \neq \surd} 
         {C;\!D \trans{\lambda} C';\!D}                  
\qquad

\bigfract{C \trans{\surd} C' \quad D \trans{\lambda} D'}
         {C;\!D \trans{\lambda} D'}        
}\\
\\
\qquad
\bigfract{C \trans{\surd} C' \quad D \trans{\surd} D'}
         {C \!\parop\! D \trans{\surd} C' \!\parop\! D'}  
\qquad
&
\bigfract{C \trans{\lambda} C' \quad \lambda \neq \surd}
         {C \!\parop\! D \trans{\lambda} C' \!\parop\! D}                           
\\
\\
C^* \trans{\surd} \pinull 
&
\bigfract{C \trans{\lambda} C' \quad \lambda \neq \surd }
         {C^* \trans{\lambda} C' ; C^*}  
\end{array}
\]

\caption{Semantic rules for orchestrations (symmetric rules omitted).}
\label{contsem}

\end{table}

\begin{table}[tbh]
\[
\begin{array}{clclc}
\bigfract{C \trans{a} C' \quad }
         {[C]_r \trans{a_r} [C']_r} 
& \qquad &
\bigfract{C \trans{\overline{a}_{s}} C' \quad }
         {[C]_r \trans{\overline{a}_{rs}} [C']_r} 
& \qquad &
\bigfract{P \trans{\lambda} P' \quad \lambda \neq \surd }
         {P  \pa Q \trans{\lambda} P' \pa Q} 
\\
\\
\multicolumn{5}{c}{
\bigfract{P \trans{a_s} P' \quad
          Q \trans{\overline a_{rs}} Q'}
         {P \pa Q \trans{a_{r\rightarrow s}} P' \pa Q'} 
\qquad
\bigfract{P \trans{\surd} P' \quad
Q \trans{\surd} Q'}
{P \pa Q \trans{\surd} P' \pa Q'}
}
\\
\end{array}
\]

\caption{Semantic rules for orchestration compositions (symmetric rules omitted).}
\label{contcompsem}

\end{table}

Here and in the remainder of the paper we use the following
notations: 
$P \trans{\lambda}$ to mean that there exists $P'$ 
such that $P \trans{\lambda} P'$ and,
given a sequence of labels $w=\lambda_1 \lambda_2 \cdots \lambda_{n-1}\lambda_n$
(possibly empty, i.e., $w = \vuota$), 
we use $P \trans{w} P'$ to denote the sequence of
transitions $P \trans{\lambda_1} P_1 \trans{\lambda_2}
\cdots \trans{\lambda_{n-1}} P_{n-1} \trans{\lambda_n} P'$
(in case of $w=\vuota$ we have $P'=P$, i.e., $P \trans{\vuota} P$).
Moreover, for completely specified systems $P$ (i.e. terms $P$ not included as subterms in a larger term $P'$), we do not consider transitions corresponding to unmatched input and output actions: namely,
we consider only transitions labeled with $\tau$ (local internal
actions), $\surd$ (global successful termination) and 
$a_{r\rightarrow s}$ (completed interactions).

We now define the notion of correct
composition of orchestrations. This notion is the same as in~\cite{fsen07}. Intuitively, a system
composed of orchestrations is correct if all possible
computations may guarantee completion; this 
means that the system is both deadlock and livelock free
(there can be an infinite computation, but given
any possible prefix of this infinite computation, it must be possible to extend it 
to reach a successfully completed computation).

\begin{definition}\label{correctcomposition}
{\bf (Correct orchestration composition)}
System $P \in {\cal P}$ is a correct orchestration composition, 
denoted $P\!\downarrow$,
if for every
$P'$ such that
$
P \; \trans{w} \; P' 
$
there exists $P''$ such that 
$
P' \; \trans{w'} \; P'' \; \trans{\surd}. 
$
\end{definition}

\subsection{Choreography Implementation, Projection and Well-Formedness}

We are now ready to formalize the notion of correct implementation 
of a choreography.
With $P \trans{\tau^*} P'$ we denote the existence
of a (possibly empty) sequence of $\tau$-labeled transitions 
starting from the system $P$ and leading to $P'$.
Given the sequence of labels $w = \lambda_1 \cdots \lambda_n$,
we write $P \stackrel{w}{\Longrightarrow} P'$ if 
there exist $P_1, \cdots, P_m$ such that 
$P \trans{\tau^*} P_1 \trans{\lambda_1} P_2 \trans{\tau^*} \cdots \trans{\tau^*}
 P_{m-1} \trans{\lambda_n} P_m \trans{\tau^*} P'$.

Intuitively, a system implements a choreography if it 
is a correct composition of orchestrations and 
all of its conversations  (i.e. 
the possible sequences of message exchanges), 
are admitted by the choreography.

\begin{definition}\label{impl}{\bf (Choreography implementation)}
Given the choreography $H$ and the system $P$, we say that
$P$ implements $H$ (written $P \propto H$) if
\begin{itemize} 
\item
$P$ is a correct orchestration composition and
\item
given a sequence $w$ of labels of the kind $a_{r \rightarrow s}$,
if $P \stackrel{w \surd}{\Longrightarrow} P'$
then there exists $H'$ such that $H \stackrel{w \surd}{\longrightarrow} H'$.
\end{itemize}
\end{definition}

Note that it is not necessary for an implementation
to include all possible conversations admitted
by a choreography.

\begin{example}{\bf (Implementation of Buyer/Seller/Bank)}
As an example, we present a possible implementation
of the choreography reported in the Example~\ref{BSB}.
\[
\begin{array}{l}
\ [\overline{Request}_{Seller} ; Offer ;
\overline{Payment}_{Bank} ; Receipt
]_{Buyer}\ 
\pa \\
\ [Request ; (\overline{Offer}_{Buyer} \parop                \overline{PayDescr}_{Bank});Confirm]_{Seller} 
\ \pa \\
\ [PayDescr ; Payment; (\overline{Receipt}_{Buyer} \parop  \overline{Confirm}_{Seller})]_{Bank}
\end{array}
\]

\end{example}

We now present the notion of choreography projection, which yields an orchestration $C$ for each role of a choreography $H$.
The definition is very simple thanks to the idea, we introduced in \cite{SC07}, of projecting communication atoms and then applying homomorphism over all the algebra operators.

\begin{definition}{\bf (Choreography projection)} 
\label{def:proj}
Given a choreography $H$, the projection 
$H$ on the role $r$, denoted with $\sem{H}_r$, is defined 
inductively on the syntax of $H$ in such a way that
\[
\sem{a_{r \rightarrow s}}_t\ =\
\left \{
\begin{array}{ll}
\overline{a}_s \qquad  & \mbox{if $t=r$} \\
a                           & \mbox{if $t=s$} \\ 
\pisucc                     & \mbox{otherwise}
\end{array}
\right .
\]
and that it is a homomorphism with respect to all
operators.
\end{definition}

It is interesting to observe that given a choreography $H$,
the system obtained composing its 
projections is not ensured to be an implementation of $H$.
For instance, consider the choreography
$
a_{r \rightarrow s}\ ;\ b_{t \rightarrow u} 
$.
The system obtained by projection is 
$
[\overline{a}_s]_r\ \pa\ [a]_s\ \pa\ [\overline{b}_u]_t\ \pa\ [b]_u
$.
Even if this is a correct composition of orchestrations,
it is not an implementation of $H$ because it comprises
the conversation $b_{t \rightarrow u} a_{r \rightarrow s}$ which
is not admitted by $H$.

The problem is not in the definition of the projection,
but in the fact that the above choreography
cannot be implemented preserving the 
message exchanges specified by the choreography. 
In fact, in order to guarantee that
the communication between $t$ and $u$ is executed after the
communication between $r$ and $s$, it is necessary to 
add a further message exchange (for instance between
$s$ and $r$) which is not considered in the choreography.


In order to have the guarantee that the system obtained
by projection is consistent with the initial choreography,
it is reasonable to consider a subset of {\em well formed}
choreographies.
The most general and intuitive notion of well formedness, we introduced in \cite{SC07},
can be obtained by
considering only all those
choreographies for which the system obtained by
projection is ensured to be a correct implementation.
\begin{definition}\label{wellFormed}{\bf (Well formed choreography)}
A choreography $H$, defined on the roles $r_1,\cdots,r_n$,
is {\em well formed} if 
$
[\, \sem{H}_{r_1}\, ]_{r_1}\ \pa\ \cdots\ \pa\ [\, \sem{H}_{r_n}\,]_{r_n}
\propto\ H
$
\end{definition}
It is worthwhile to note that well formedness is
decidable. In fact, given a choreography $H$,
it is sufficient to take the corresponding 
system $P$ obtained by 
projection, then consider $P$ and $H$ as finite 
state automata, and finally check whether the language 
of the first automaton is included in the language of the 
second one.
Note that the terms $P$ and $H$ can be seen as finite
state automata
thanks to the fact that their infinite behaviours
are defined using Kleene-star repetitions instead 
of general recursion.

This decidability result clearly follows from the fact that
we restrict to finite state choreographies. In the literature, 
more general syntactic versions of well formedness exist
(see e.g.~\cite{CHY07}). In the next section we report a syntactic
characterisation of well formedness.

\subsection{Connected Choreographies}
\label{sec:connect}

In this section, following~\cite{sefm08,tgc08}, we introduce a notion of connectedness
for our choreography calculus. The idea is to impose syntactic
restrictions in order to avoid the three possible ways in which
a system obtained by projection can have a different behaviour
w.r.t. its choreography.

The first of these ways is concerned with sequential composition: given 
the choreography $H_1;H_2$, if in the projected system there is
no synchronisation between the roles involved in $H_1$ and
those in $H_2$, an interaction in the latter could occur before
an interaction in the former.
This first problem can be avoided by imposing that for every
initial interaction in $H_2$ there is at least one role involved
in every possible final interaction of $H_1$.
To formalise this syntactic restriction we use two auxiliary functions
$\transI(\bullet)$ and $\transF(\bullet)$ that computes respectively the set of 
pair of roles involved in initial and
final interactions in a choreography:
\[
\begin{array}{l}
\transI(\chor a{r_1}{r_2}) = \transF(\chor a{r_1}{r_2}) = \big\{\{r_1,r_2\}\big\} \\
\transI(\pisucc)=\transI(\pinull)=\transF(\pisucc)=\transF(\pinull)=\emptyset\\
\transI(H \!\parop\! H') = \transI(H + H') = \transI(H) \cup \transI(H')\\
\transF(H \!\parop\! H') = \transF(H) \cup \transF(H')\\
\transF(H + H') = \transF(H) \cup \transF(H')\\
\transI(H;H')  = \transI(H) \cup \transI(H') \mbox{if $H \trans{\surd}$, $\transI(H)$ otherwise}\\
\transF(H;H')  = \transF(H) \cup \transF(H') \mbox{if $H' \trans{\surd}$, $\transF(H')$ otherwise}\\
\transI(H^*)=\transI(H)\\
\transF(H^*)=\transF(H)\\
\end{array}
\]
The following syntactic condition on choreographies guarantees
that the above problem does not occur.

\begin{definition}\label{connectedSequence}
{\bf (Connectedness for sequence)}
A choreography is connected for sequences if for each
subterm of the form $H;H'$ we have that $\forall R \in \transF(H),
\forall R' \in \transI(H')$ then $R \cap R' \neq \emptyset$.
\end{definition}

The second way a projected system can differ from its choreography
is concerned with the choice operator: in a choreography $H+H'$ it is
necessary that all the involved roles are aware of the selected branch.
This can be ensured by imposing that at least one of the roles
involved in initial interactions in one of the two branches
is involved in every initial interaction in the other one.
Moreover, it is also necessary to impose that the two choreographies
consider the same set of roles: this guarantees 
that precisely the same set of roles are involved in the two
branches thus, even if not initiator, they are eventually informed
of the selected branch.

\begin{definition}\label{connectedChoice}
{\bf (Unique point of choice)}
A choreography has unique points of choice if for
each subterm of the form $H+H'$ we have that $\forall R \in \transI(H),
\forall R' \in \transI(H')$ then $R \cap R' \neq \emptyset$
and moreover $roles(H)=roles(H')$.
\end{definition}

Finally, the last way projections can differ from choreographies
is concerned with parallel composition. In the projection of 
a choreography $H|H'$, it could happen that a message sent within
an interaction in one choreography is 
intercepted by a receive action in an interaction
of the other one.
We can avoid this by imposing that the interactions in $H$ 
uses names different from those used in $H'$.

\begin{definition}\label{connectedParallel}
{\bf (No operation interference)}
A choreography has no operation interferences
if for every pair of interactions  
$\chor a{r_1}{r_2}$
and $\chor {a'}{r'_1}{r'_2} $ occurring in parallel
in the same scope, we have $a \neq a'$.
\end{definition}

The first two conditions have been borrowed from~\cite{sefm08},
while the last one is a simpler and stronger condition w.r.t. 
the (more complex but less restrictive) {\em causality safety} 
condition considered in that paper. One of the main
results in~\cite{sefm08} is that when choreographies respect 
the three connectedness conditions, then they are bisimilar
to their projections. We can then conclude what follows.

\begin{theorem}
Let $H$ be a choreography satisfying the 
conditions in the Definitions~\ref{connectedSequence},~\ref{connectedChoice}, and~\ref{connectedParallel}.
Then $H$ is {\em well formed}.	
\end{theorem}

\section{Contract-based Service Discovery}


%

We now define the notion of behavioural contract which will allow us to reason about service compliance and 
retrieval independently of the language used for implementing service behaviour. 

Contracts are defined as labeled transition systems, where transition labels are the typical internal $\tau$ action and the
input/output actions $a,\overline{a}_l$, where the outputs (as for the orchestration calculus) are directed to a destination address denoted by a role $r \in Roles$.
We first define the class of labeled transition systems
of interest for defining contracts.

\begin{definition} {\bf (Finite Connected LTS with Termination Transitions)}
A finite connected labeled transition system (LTS) with termination transitions is a tuple ${\cal t} = (S,{\cal L},\trans{},s_h,s_0)$ where $S$ is a finite set of states, 
$L$ is a set of labels, the transition relation $\trans{}$ is a finite subset of $(S-\{s_h\})
 \times ({\cal L} \cup \{ \surd \}) \times S$ such that $(s, \surd, s') \in \trans{}$ implies $s' = s_h$, $s_h \in S$ represents a halt state, $s_0 \in S$ represents the initial state,
 and it holds that every state in $S$ is reachable (according to $\trans{}$) from $s_0$.
\end{definition}

As in orchestrations, in a finite connected LTS with termination transitions
we use $\surd$ transitions (leading to the halt state $s_h$) to represent successful termination. On the contrary, if we get 
(via a transition different from $\surd$) into a state with no outgoing transitions (like, e.g., $s_h$) 
then we represent an internal failure or a deadlock.


\begin{definition}{\bf (Behavioural Contract)}\label{contractdef}
\label{def:contract}
A behavioural contract is a finite connected LTS with termination transitions,
that is a tuple $(S,{\cal L},\trans{},s_h,s_0)$, where ${\cal L} = \{a,\overline{a}_l,\tau
\ |\ a \in {\cal N} \wedge l \in Roles \}$, 
i.e. labels are either a receive (input) on some operation $a \in {\cal N}$ or an invoke (output) directed to some operation $a \in {\cal N}$ at some role $l$ or internal $\tau$ actions.
\end{definition}


Orchestrations $C \in {\cal P}_{orc}$ give rise to a behavioral contract as the labeled transition system
obtained by their semantics (where the initial state $s_0$ is $C$, the halt state $s_h$ is $\pinull$ and the other states in $S$ are terms $C'$ reachable by $C$).
We can, therefore, use $C \in {\cal P}_{orc}$ as a way to denote such a behavioral contract.

In the following we will present the theory of behavioral contracts by abstracting from the particular process algebra used to denote them: contracts represent orchestration behavior in a {\em language independent way}. 
We will therefore take terms $C$ belonging to a generic set
${\cal P}_{con}$ being any set of terms (generated by the syntax of a process algebra) that give rise to all possible behavioral contracts as their semantics (in the form of finite connected LTSes as in Definition \ref{contractdef}). 
For instance ${\cal P}_{con}$ can be basic CCS  (with recursion) over ${\cal L}$ prefixes and extended with successful termination $\pisucc$, see \cite{libroCubo,sensoria}.

\subsection{Contract Compositions and Contract Compliance}

Similarly as for orchestrations, we consider {\it contract compositions} $P$ as parallel compositions of contracts $C$, where the syntax and semantics of $P$ is defined as for the syntax (Definition \ref{orcsys}) and semantics (Table \ref{contcompsem}) of systems, where we now take $C$ belonging to 
the arbitrary
 set ${\cal P}_{con}$, instead of $C \in {\cal P}_{orc}$. With a little abuse of notation in this section we will use ${\cal P}$ to denote the set of such contract compositions $P$ (the same notation used for set of systems in the previous section).

When reasoning about contract compositions $P$, it will be fundamental to consider (as we did for the particular orchestration language of the previous section) correct contract compositions, denoted by $P\downarrow$.
The notion of {\it correct contract composition} (also called {\it contract compliance}) $P$ is defined exactly as for correct orchestration composition, see Definition \ref{correctcomposition} (the only difference now being that $P$ is composed of terms $C$ belonging to the arbitrary set ${\cal P}_{con}$, instead of $C \in {\cal P}_{orc}$).

\subsection{Independent Subcontracts}

\noindent
We are now ready to define the notion of independent subcontract
pre-order. Given a contract $C \in {\cal P}_{con}$, we use 
$oroles(C)$ to denote the subset of $Roles$ of
the roles of the destinations of all the output actions
occurring inside $C$.

\begin{definition}\label{defsubfam} {\bf (Independent Subcontract pre-order)}
A pre-order $\leq$ over ${\cal P}_{con}$ is an independent subcontract pre-order if, 
for any $n \geq 1$, contracts $C_1, \dots, 
C_n \in {\cal P}_{con}$ and $C_1', \dots, C_n' \in {\cal P}_{con}$
such that $\forall i \ldotp C_i' \leq C_i$,
and distinguished role names $l_1, \dots, l_n \in {Roles}$ such that 
$\forall i \ldotp l_i \notin oroles(C_i) \cup oroles(C_i')$, we have
\[([C_1]_{l_1} \; \pa  \dots  \pa  \; 
[C_n]_{l_n})\!\downarrow  \quad \Rightarrow 
\quad ([C_1']_{l_1} \; \pa  \dots  \pa \; [C_n']_{l_n})\!\downarrow
\]
\end{definition}

It is easy to see that, if we do not introduce any asymmetry in the behaviour of invokes and receives, there is no maximal independent subcontract pre-order, i.e. there is no optimal solution to the problem of locally retrieving services based 
on their contracts.

This can be easily seen by considering, e.g., the trivially 
correct system
$[C_1]_l \pa [C_2]_{l'}$ with $C_1 = a$ and $C_2 = \overline{a}_l$. 
Consider the two independent subcontract pre-orders 
$\leq^1$ and $\leq^2$ such that the unique pairs they possess besides the identity on all contracts $C$ are
\[
a+c;\pinull \leq^1 a\
\]
and
\[
\overline{a}_l+\overline{c}_{l};\pinull \leq^2 \overline{a}_l
\]
No pre-order $\leq$ could have both
\[
a+c;\pinull  \leq a\
\mbox{\ and\ }\ 
\overline{a}_l+\overline{c}_{l};\pinull  \leq \overline{a}_l
\]
because if we refine $C_1$ with $a+c;\pinull$ and 
$C_2$ with $\overline{a}_l+\overline{c}_l;\pinull$ we achieve the 
incorrect system 
$a+c.\pinull \pa \overline{a}_l+\overline{c}_l;\pinull$
that can deadlock after synchronization on channel $c$.

\noindent

We now show that, by assuming a more practical form of communication,
where invokes and receives have an asymmetric behaviour, 
there exists a maximal independent subcontract pre-order.
This con be achieved in several ways:
$(i)$ constraining the structure of behavioural contracts by considering only {\it output persistent} ones (a contract chooses internally to perform an output/invoke: it must execute it to reach successful termination) as in WS-BPEL~\cite{wsbpel-spec} or in session types~\cite{CHY07};
$(ii)$ strengthening the notion of compliance (yielding so-called strong compliance \cite{mscs09,coord07}): when an output is performed a corresponding input is required to be already enabled, like in ready-send of Message Passing Interface 
$(iii)$ moving to asynchronous communication, e.g. via message queues \cite{libroCubo,wsfm08}.

In the following we will detail the first of the above approaches. In Section \ref{summary} we will survay other approaches.

\subsection{Output Persistence}

The main results reported in this paper are
consequences of a property of systems that
we call {\em output persistence}.




\begin{definition}\label{outper}{\bf(Output persistence)}
Let $C \in {\cal P}_{con}$ be a behavioural contract.
It is output persistent if given
$C \trans{w} C'$ with $C' \trans{\overline{a}_l}$ then:
$C' \ntrans{\surd}$ and
if $C' \trans{\alpha} C''$ with $\alpha \neq \overline{a}$ 
then also $C'' \trans{\overline{a}_l}$.
\end{definition}


The output persistence property states that once a contract 
decides to execute an output, 
its actual execution is mandatory in order to 
successfully complete the execution of the contract.
This property typically hold in languages for the
description of service behaviours or for
service orchestrations (WS-BPEL~\cite{wsbpel-spec} or session types~\cite{CHY07})
in which output actions cannot be used as guards
in external choices (see e.g. the {\tt pick}
operator of WS-BPEL which is an external choice
guarded on input actions). In these languages,
a process instance always  decides {\it internally} to execute 
an output action (output action cannot be involved in an external choice).
In fact, if we consider potential outputs that can disappear
without being actually executed (as in an external choice among outputs $\overline{a}+\overline{b}$ or in a mixed choice $a+\overline{b}$
in which, e.g., the possible $\overline{b}$ is no longer executable
after the output or input on $a$) we have that
there exists no maximal subcontract
pre-order family, as we have shown with the above counterexample. 

Notice that, if we instead assume output persistence of contracts, as we do in this paper, 
subcontracts cannot add reachable outputs on new types. For instance 
an output persistent contract $a+\tau.\overline{c_l}$ adding a new output on $c_l$
with respect to $a$, similarly to the pre-order 
$\leq^2$ in the counterexample above, would not be a correct subcontract because when composed in parallel with 
the other initial contract $\overline{a}_l$ would lead to a deadlock.

In the context of process algebra with parallel composition as that of ${\cal P}_{orc}$, a syntactical characterization
that guarantees output persistence has been introduced in~\cite{fsen07}.
The idea is to require that every output prefix (i.e. the term
$\overline{a}_l$) is preceded by an internal $\tau$ prefix, i.e. the syntax of Definition \ref{orcsys} includes
$\tau;\overline{a}_l$ instead of $\overline{a}_l$.
Choreography projection must be also modified in order to produce, for each role, an output persistent contract that can be then used for discovering services via refinement (see following Section \ref{chorconf}). 
In particular, we have to modify the rule concerning generation of output in orchestrations as follows:
$$\sem{a_{r \rightarrow s}}_t\ =\ \tau;\overline{a}_s \qquad \mbox{if $t=r$}$$

In the following we will take ${\cal P}_{opcon}$ to be any set of terms (generated by the syntax of a process algebra) that give rise to all possible output persistent behavioral contracts as their semantics (in the form of finite connected LTSes as in Definition \ref{contractdef}). Moreover we will consider systems ${\cal P}$ to be compositions of contracts in
${\cal P}_{opcon}$.

\subsection{Compliance Testing is the Maximal Preorder}\label{subrel}

We will show that the maximal independent subcontract
pre-order can be achieved defining a
more coarse form of refinement in which,
given any system composed of a set of contracts,
refinement is applied to one contract only
(thus leaving the others unchanged). This
form of refinement, that we call {\em compliance testing}, is a form of testing
where both the test and the system under test must reach success.

Given a system $P \in {\cal P}$, we use $roles(P)$ to denote the subset of
$Roles$ of the roles of contracts syntactically occurring inside $P$:
e.g. $roles([C]_{l_1} \pa [C']_{l_2}) = \{ l_1, l_2 \}$.

\begin{definition} {\bf (Subcontract Relation)} \label{defsubrel}
A contract $C' \in {\cal P}_{opcon}$ is a subcontract of a contract $C \in {\cal P}_{opcon}$ 
denoted $C' \preceq C$,
if and only if for all $l \in {Roles}$ such that $l \notin oroles(C_i) \cup oroles(C_i')$ and
$P \in {\cal P}$ such that $l \notin roles(P)$ we have
$
([C]_l \pa P)\!\downarrow 
$ implies $
([C']_l \pa P)\!\downarrow
$
\end{definition}

\begin{theorem}\label{hard}
There exists a maximal independent subcontract
pre-order and it corresponds to the (compliance testing based) subcontract relation ``$\preceq$''. Formally, for any pre-order $\leq$ over ${\cal P}_{opcon}$ that is an independent subcontract pre-order, we have that $\leq$ is included in $\preceq$.
\end{theorem}

%
%
%
%




\subsection{Properties of the Maximal Preorder and I/O Knowledge}

We now discuss some properties of the maximal independent subcontract pre-order (compliance testing), comparing it with other approaches and other
classical preorders. We also show a sound characterization that is decidable.

In the following we will use $I(C)$ to stand for the subset of ${\cal N}$ of the input actions $a$ syntactically occurring in $C$ and ``$C \re M$'', with $M \subseteq {\cal N}$, to stand for ``$C \{\pinull / a | a \in M \}$''.

\begin{proposition}\label{lemmuccio}
Let $C,C' \in {\cal P}_{opcon}$ be contracts. 
We have
\[
C' \re (I(C') - I(C)) \preceq C \quad \Leftrightarrow \quad C'  \; \preceq C
\]
\end{proposition}

The proposition above is a direct consequence of the fact that, due to output persistence of contracts,  
compliant tests $P$ of a contract $[C]_l$ cannot perform reachable outputs directed to $l$ that $C$ cannot receive.
For instance $[a]_l \pa [\tau;\overline{a}_l + \tau;\overline{b}_l]_{l'}$ is not a correct contract composition.

From the proposition above we can derive two fundamental properties of the maximal independent subcontract pre-order:
\begin{itemize}
\item External choices on inputs can be extended, for instance:
$$a+b \preceq  a$$
by directly applying the above proposition.
\item Internal choices on outputs can be reduced, for instance:
$$\tau;\overline{a}_l  \preceq   \tau;\overline{a}_l + \tau;\overline{b}_l$$
because the lefthand term is more deterministic (typical property in testing).
\end{itemize}
These two properties are assumed when considering subtyping in the theory of session types \cite{CHY07}, in our setting
they were instead obtained as a consequence of considering the maximal independent subcontract preorder over output persistent contracts.

Notice that, when considering behavioural contracts that communicate just with standard CCS
channel name based communication \cite{sfm09,fundaInfo08,fsen07}, i.e. where outputs are just written as $\overline{a}$ instead of being directed to a certain role $l$ as in $\overline{a}_l$, the proposition above does not hold.
This is due to ``capturing'' behaviour. For instance it does not hold that $a+b \preceq  a$, as can be seen by considering the correct contract composition 
$$[a] \pa [\tau;\overline{a};b ] \pa [\tau;\overline{b}]$$
and the incorrect one
$$[a+b] \pa [\tau;\overline{a};b ] \pa [\tau;\overline{b}]$$
where the leftmost contract can ``capture'' the $\overline{b}$ that was received by the middle contract in the correct composition.

This problem can be faced by resorting to knowledge about the input and output actions syntactically occurring in the initial 
contracts (e.g. those obtained by projection from a choreography) when defining the notion of independent subcontract preorder $\leq$, see \cite{sfm09,fundaInfo08,fsen07}. In this way $C' \preceq_{I,O} C$ denotes that $C'$ is a subcontract of $C$ assuming that the other contracts (the test) can only perform input on $I$ and output on $O$.
Therefore, exploiting knowledge we can recover the above proposition and we have:
$$a+b \preceq_{{\cal N},{\cal N}-\{ b \}}  a$$

\subsubsection{Resorting to Fair Testing}\label{should}

This section is devoted to 
the definition of an actual procedure for determining that
two contracts are in subcontract relation.
This is achieved resorting to the theory of fair testing, called
{\em should-testing}~\cite{RV05}. As a side effect we will also show that subcontract relation is coarser than fair testing preorder.


In the following we denote with $\preceq_{test}$ the {\em should-testing} pre-order
defined in~\cite{RV05} where we consider the set of actions
used by terms as being ${\cal L} \cup \{ \overline{a} \mid a \in {\cal N} \}$ 
(i.e. we consider located and unlocated input and output actions and $\surd$ is included
in the set of actions of terms under testing as any other
action). 
We denote here with $\surd'$ the special action
for the success of the test (denoted by $\surd$ in~\cite{RV05}).
In the following we consider $\lambda$ to range over ${\cal L} \cup \{ 
\overline{a} \mid a \in {\cal N} \}$.

In order to resort to the theory defined in~\cite{RV05},
we define a normal form for contracts
of our calculus that corresponds to 
terms of the language in~\cite{RV05}.
The normal form of the contract $C$ (denoted
with ${\cal NF}(C)$) is defined as follows,
by using the operator $rec_{X} \theta$ 
(defined in~\cite{RV05}) that represents the value of $X$ 
in the solution of the minimum fixpoint of the finite
set of equations $\theta$,
\[
\begin{array}{rll}
{\cal NF}(C) & = & rec_{X_1} \theta \qquad\mbox{where $\theta$ is the set of equations}
\\
X_i & = & \sum_{j} \lambda_{i,j} ; X_{der(i,j)}
\end{array}   
\]
where, assuming to enumerate the states
in the labeled transition system of $P$
starting from $X_1$,
each variable $X_i$ corresponds to 
the $i$-th state of the labeled transition system
of $P$, $\lambda_{i,j}$ is the label of the $j$-th
outgoing transition from $X_i$, and ${der(i,j)}$
is the index of the state reached with 
the $j$-th outgoing transition from $X_i$.
We assume empty sums to be equal to $\pinull$,
i.e. if there are no outgoing transitions
from $X_i$, we have $X_i = \pinull$.

\begin{theorem}\label{finale}
Let $C,C' \in {\cal P}_{opcon}$ be two contracts.
We have
\[
{\cal NF}(C' \re\!I(C')\!-\!I(C)) \preceq_{test} {\cal NF}(C) \quad \Rightarrow
\quad C' \preceq C
\]
\end{theorem}

The opposite implication 
$
C' \preceq C
 \Rightarrow 
{\cal NF}(C' \re\! I(C')\!-\!I(C)) \preceq_{test} {\cal NF}(C) 
$
does not hold in general. 
This can be easly seen by considering uncontrollable contracts, i.e. contracts for which there is no compliant test.
For instance the contract $\pinull$, any other contract $a;b;\pinull$ or $c;d;\pinull$ or more complex examples like 
$a.\pisucc+a.b.\pisucc$.
These contracts are all equivalent according to our subcontract relation, but of course not according to fair testing.

\subsubsection{Comparison with Traditional Testing and Trace Preorders}

Notice that Theorem \ref{finale} says that the subcontract relation is coarser than fair testing preorder (where
tests are assumed to be capable of observing $\surd$ actions of the system under test).
The equivalent uncontrollable contracts that we have shown in the previous section show that this inclusion
is strict: this is due to the fact that, besides requiring the success of the test,
we impose also that the tested process should
successfully complete its execution. 

Another significant difference of compliance testing respect to traditional
one proposed by De Nicola-Hennessy~\cite{DH84}
 is in the treatment of divergence: we do not
follow the traditional catastrophic approach,
but the fair approach as in the theory
of should-testing of Rensink-Vogler~\cite{RV05}. 
In fact, we do not
impose that all computations must succeed, but that
all computations can always be extended in order
to reach success.

Concerning trace preorder, it is not coarser than subcontract relation (compliance testing) due
to the possible presence in the system under test of traces not leading to success: these traces are not observable by (compliant) tests. For instance the equivalent uncontrollable contracts that we have shown in the previous section 
have completely different traces. In this respect compliance testing differs from standard testing theories which imply
trace inclusion.

\subsection{Contract-based Choreography Conformance}\label{chorconf}

We are now in place for the definition
of relations $C \triangleleft_r H$
indicating whether the contract $C$ can
play the role $r$ in the choreography $H$.

\begin{definition}{\bf (Conformance relation)}
Given a well-formed choreography
$H$
with roles $r_1, \cdots, r_n$,
a relation among contracts, roles $r$ of $H$ and $H$,
denoted  with 
$C \triangleleft_{r} H$,
is a {\it conformance relation} if, for every $r_i$, with 
$1 \leq i \leq n$, there exists at least a contract $C_i$ such that
$C_i \triangleleft_{r_i} H$, and it holds that:
$
\mbox{if}\  
C_1 \triangleleft_{r_1} H , \cdots, C_n \triangleleft_{r_n} H 
\ \mbox{then}\ 
[C_1]_{r_1} \pa \cdots \pa [C_n]_{r_n} 
\propto H
$.
\end{definition}

It is interesting to observe that, differently
from subcontract relation
defined on contracts in the previous Section \ref{subrel},
there exists no maximal conformance relation.
For instance, consider the choreography 
$H = a_{r \rightarrow s}|b_{r \rightarrow s}$.
We could have two different conformance relations $\triangleleft^1$ and $\triangleleft^2$,
the first one such that 
\[
(\tau;\overline{a}_s | \tau;\overline{b}_s) \triangleleft^1_r H
\qquad \qquad
(\tau;a;b\ +\ \tau;b;a) \triangleleft^1_s H
\]
and the second one such that
\[
(\tau;\overline{a}_s;\tau;\overline{b}_s + \tau;\overline{b}_s;\tau;\overline{a}_s) 
\triangleleft^2_r H
\qquad \qquad
(a | b) \triangleleft^2_s H
\]
It is easy to see that it is not possible to have 
a conformance family that comprises the union of the
two relations $\triangleleft^1$
and $\triangleleft^2$. In fact, the system
\[
[\tau;\overline{a}_s;\tau;\overline{b}_s\ +\ \tau;\overline{b}_s;\tau;\overline{a}_s]_r\ \pa\
[\tau;a;b\ +\ \tau;b;a]_s
\]
is not a correct composition because
the two contracts may internally select two 
incompatible orderings for the execution
of the two message exchanges (and in this
case they stuck).

In the remainder of this section we define
a mechanism that, exploiting the notion of subcontract relation defined in the previous section and, in particular, its sound characterization based on fair testing, 
permits to effectively characterize an interesting conformance relation. 

We now present  the notion of {\em testing consonance} (see \cite{libroCubo})
between contracts and roles of a given choreography,
and we prove that it is a conformance family.
\begin{definition}{\bf (Testing Consonance)}
We say that the contract $C$ is {\em testing consonant}
with the role $r$ of the well formed choreography $H$
(written $C \otimes^r_{test} H$) if 
$${\cal NF}\big(  C \re  I(C) - I(\sem{H}_r) \big) \preceq_{test} 
 {\cal NF}\big( \sem{H}_r \big)$$
\end{definition}

\begin{theorem}
Given a well-formed choreography $H$, we have that the testing consonance relation $C \otimes^r_{test} H$
is a conformance relation.
\end{theorem}


\subsection{Summary of Results}\label{summary}


We now summarize the results that we have obtained 
about 
contract refinement and choreography conformance
for different forms of communication.

The first mean of classification of possible scenarios is based on the {\it amount of knowledge} about the (initial) behavioural description of the other roles the conformance relation may depend on. We considered: knowledge about the
whole choreography (full knowledge about the behaviour of other roles) or knowledge restricted to input types (receive operations) and/or output types (invoke operations) that the other roles may use. 
The second mean of classification of possible scenarios is based on the {\it kind of service compliance} assumed (i.e. of the principle assumed for assessing when multiple services work well together and form a correct system). We considered: $(i)$ ``normal'' compliance, as reported in this paper, where service interaction via invoke and receive primitives is based on synchronous handshake communication and both receive and invoke primitives may wait indefinitely (with no exception occurring) for a communication to happen, see \cite{sfm09,fundaInfo08,fsen07} and \cite{libroCubo,sensoria,SC07} where, in addition to the standard CCS synchronization, locations expressing a unique address for every system contract are introduced and outputs are directed to a location (as in the thoery presented in this paper);
$(ii)$ ``strong compliance'',
where we additionally require that, whenever a service may perform an invoke on some operation, the invoked service must be already in the receive state for that operation as for the ready-send primitive of the Message Passing Interface \cite{mscs09,coord07}; 
$(iii)$ ``queue-based compliance'',
where service interaction via invoke and receive primitives
is based on asynchronous communication: the receiving service puts
invoke requests in an unbounded queue \cite{libroCubo,wsfm08}. 

Concerning service compliance
we considered in all cases the fair termination property presented in this paper.
Our results are summarized in the following.
\begin{itemize}
\item Knowledge about the whole choreography (direct conformance relation with respect to a choreography for a certain role): the maximal independent conformance relation does not exist, no matter which
kind of service compliance (among those mentioned above) is considered.
\item Knowledge about other initial contracts limited to input/output types they use
(conformance by means of refinement of a single contract parameterized on the I/O knowledge about the others, as in this paper):
\begin{itemize}
\item In the case of ``normal compliance'' we have that: for unconstrained contracts the maximal independent conformance relation does not exist; for contracts such that the output persistence property holds the maximal independent conformance relation exists and knowledge about input types is not significant; for output persistent contracts where locations expressing a unique address for every system contract are introduced and outputs are directed to a location, as for the theory in this paper, the maximal relation exists and knowledge about both input and output types is not significant.
%
\item In the case of ``strong compliance'' we have that: for unconstrained contracts (where outputs are directed to a location identifying a unique system contract) the maximal relation exists and knowledge about both input and output types is not significant.
\item In the case of ``queue-based compliance'' we have that: for unconstrained contracts (where outputs are directed to a location identifying a unique system contract) the maximal relation exists and knowledge about both input and output types is not significant.
\end{itemize}
\end{itemize}
For every maximal refinement relation above (apart from the queue-based one), we provide
a sound characterization that is decidable, by resorting to 
an encoding into should-testing~\cite{RV05}.
As a consequence we obtain:
\begin{itemize}
\item An algorithm (based on that in~\cite{RV05}) to check
refinement.
\item
A classification of the maximal refinement relations with respect to existing pre-orders as, e.g., (half) bisimulation, (fair/must) testing, trace inclusion. 

In particular we show that the maximal refinement relations are coarser with respect to bisimulation and must testing preorders (up to some adequate encoding and treatment of fairness) in that, e.g., they allow external non-determinism on inputs to be added in refinements.
\end{itemize}

\section{Towards Choreographies and Orchestrations with Dynamic Updates}


\newcommand{\Rule}[2]{\displaystyle\frac{#1}{#2}}

\newcommand{\arr}[1]{\xrightarrow{#1}}

\newcommand{\lts}[1]{\arr{#1}}

\newcommand{\updIndex}[3]{{#1} \{ {#2}:\;{#3}\}}
\newcommand{\upd}[3]{\updIndex{#1}{#2}{#3}}

\newcommand{\coupd}[3]{{#1}_{#2}\{#3\}}

\newcommand{\Did}[1]{(\textsc{#1})}

\newcommand{\pp}{\;\mathbf{|}\;}

\newcommand{\ppn}{\;\mathbf{|\!|}\;}

\newcommand{\pps}{\mathbf{|\!|}}

\newcommand{\nil}{\mathbf{0}}
\newcommand{\zero}{\mathbf{0}}

\newcommand{\one}{\mathbf{1}}

\newcommand{\proj}[2]{\sem{#1}_{#2}}

\newcommand{\ol}[1]{\overline{#1}}
\newcommand{\m}[1]{\mathsf{#1}}

\newcommand{\outP}[4]{#1:\ol{#2}_{#3}\langle #4\rangle}
\newcommand{\inP}[4]{#1:{#2}_{#3}(#4)}

In this section we discuss, following the approach discussed in~\cite{beat},
a possible approach for extending the Choreography Calculus with dynamic updates.
Dynamic updates can be specified by defining \emph{scopes} that delimit  
parts of choreographies that, at runtime, may be replaced by a new 
choreography, coming from  either inside or outside the system. 
Updates coming from outside may
be decided by the user through some adaptation interface, by some
manager module, or by the environment. In contrast, updates coming from inside
represent self-adaptations, decided by a part of the system towards itself
or towards another part of the system, usually as a result of some
interaction producing unexpected values. Updates from outside and
from inside are indeed quite similar, e.g., an update decided by a
manager module may be from inside if the manager behavior is part of
the choreography term, from outside if it is not.

Formally speaking, the Choreography Calculus 
is extended 
with {\em scopes} and
{\em updates} as follows:

\begin{displaymath}
  \begin{array} {rllllllllll}
    H ::= & ... \\
    \quad | \quad & X:T[H]	 &\qquad (\text{scope}) &
    \quad | \quad & \upd{X}{r}{H}	 &\qquad (\text{update})
  \end{array}
\end{displaymath} 

\noindent
where $X$ is used to range over a set of {\em scope names} 
and $T$ is a set of roles. 
Construct $X:T[H]$ defines a scope named $X$
 currently executing choreography $H$ --- the name is needed to designate it as a target for a particular
adaptation. Type $T$ is the set
of roles (possibly) occurring in the scope. This is needed since a
given update can be applied to a scope only if it specifies how all
the involved roles are adapted. Operator $\upd{X}{r}{H}$ defines
\emph{internal updates}, i.e., updates offered by a participant of the
choreography. Here $r$ denotes the role offering the
update, $X$ is the name of the target scope, and $H$ is the new
choreography.

The operational semantics for the new Choreography Calculus with updates
is defined in~\cite{beat} only for a proper subset of well defined choreographies.\footnote{We refer
the reader to~\cite{beat} for a formal definition of this subset of well
defined choreographies, here we simply informally report the imposed limitations.}
%
%
%
For instance, conditions are added in order to guarantee that 
every update $\upd{X}{r}{H}$ injects a choreography $H$ that considers only
roles explicitly indicated in the type $T$ of the updated
part $X:T[H']$. Moreover, also syntactic restrictions are imposed
in order to guarantee that it will never occur that two distinct
scopes with the same name are contemporaneously active.
This is necessary for the following reason: when a scope is projected,
it is distributed among several independent roles running in parallel, 
and in order to avoid interferences between two distinct scopes with the
same name $X$ we assume that only one scope $X$ can be active
at a time.

The operational semantics 
is defined
by adding the rules dealing with {\em scopes} and {\em updates}
reported in Table~\ref{table:semChoreo}. The transition labels $\eta$
now include the label $\upd{X}{r}{H}$ indicating the execution
of an update action. 
In the rules, we use $H [H'/X]$ to denote the
substitution that replaces all scopes $X:T[H'']$ with name $X$
occurring in $H$ (not inside update prefixes) with $X:T[H']$.  

\begin{table}[t]
\begin{displaymath}
\begin{array}{c}
   \\[6mm]
   \Did{CommUpd} ~ \Rule {} { \upd{X}{r}{H} \lts{\upd{X}{r}{H}} \one}
    \qquad
    \Did{SeqUpd} ~ 
    \Rule {H_1\lts{\upd{X}{r}{H}} H_1'} 
    {H_1;\ H_2 \lts{\upd{X}{r}{H}} H_1';(H_2 [H/X])}
    \\[6mm]
    \Did{ParUpd} ~ 
    \Rule {H_1\lts{\upd{X}{r}{H}} H_1'} 
    {H_1\ \pp\ H_2 \lts{\upd{X}{r}{H}} H_1'\ \pp\ (H_2 [H/X]) }
    \quad
    \Did{StarUpd} ~
    \Rule {H_1\lts{\upd{X}{r}{H}} H_1'} 
    {H_1^* \lts{\upd{X}{r}{H}} H_1';\ (H_1 [H/X])^*}
    \\[6mm]
    \Did{ScopeUpd} ~
    \Rule {H_1\lts{\upd{X}{r}{H}} H_1'}  
    {X:T[H_1] \lts{\upd{X}{r}{H}} X:T[H] }
    \quad
    \Did{ScopeComm} ~
    \Rule {H_1\lts{\chor {a}{r_1}{r_2}} H_1'}  
    {X:T[H_1] \lts{\chor {a}{r_1}{r_2}} X:T[H_1'] }
    \\[6mm]
    \Did{Scope} ~ 
    \Rule {H_1\lts{\eta} H_1' }
    {X:T[H_1] \lts{\eta} X:T[H_1'] } \;\; 
\eta \neq \upd{X}{r}{H} \; \mbox{for any} \; r,H 

\end{array}
\end{displaymath}
\caption{Semantics of Choreographies with updates}\label{table:semChoreo}
\end{table}


We briefly comment 
%
the 
rules in Table~\ref{table:semChoreo}.
Rule \Did{CommUpd} makes
 an internal update available, moving the information to the
label. Updates propagate through sequence, parallel composition,
and Kleene star using rules \Did{SeqUpd}, \Did{ParUpd}, and
\Did{StarUpd}, respectively. Note that, while propagating, the update is
applied to the continuation of the sequential composition, to parallel
terms, and to the body of Kleene star. Notably, the update is
applied to both enabled and non enabled
occurrences of the desired scope. Rule~\Did{ScopeUpd} allows a scope
to update itself (provided that the names coincide), while propagating
the update to the rest of the choreography. 
Rule~\Did{Scope} allows a scope to compute.

\begin{example}\label{AdaptableBSB}
{\bf (Adaptable Buyer/Seller/Bank)}
Here, we consider a version of the Buyer/Seller/Bank example
discussed in the Example~\ref{BSB} where it is possible to update
the payment interaction between the buyer and the bank by using, 
for instance, a new version 
of the payment protocol 
according to which the buyer sends its VISA code
to the bank and the bank subsequently confirms its correctness. 
Let us consider the following choreography composed of three
roles: {\em Buyer}, {\em Seller} and {\em Bank}
\[
\begin{array}{l}
Request_{Buyer \rightarrow Seller};
(\ Offer_{Seller \rightarrow Buyer} \parop
PayDescr_{Seller \rightarrow Bank}\ ); \\
\hspace{.5cm} X\{Buyer,Bank\}[Payment_{Buyer \rightarrow Bank}];
(\ Confirm_{Bank \rightarrow Seller} \parop 
Receipt_{Bank \rightarrow Buyer}\ )
\end{array}
\]
According to the operational semantics defined above,
this choreography could, for instance, perform the initial $Request$
interaction and then receives an external
update:
$$\upd{X}{r}{VISAcode_{Buyer \rightarrow Bank};VISAok_{Bank \rightarrow Buyer}}$$ 
and then becomes the following choreography:
\[
\begin{array}{l}
(\ Offer_{Seller \rightarrow Buyer} \parop
PayDescr_{Seller \rightarrow Bank}\ ); \\
\hspace{.5cm} X\{Buyer,Bank\}[VISAcode_{Buyer \rightarrow Bank};VISAok_{Bank \rightarrow Buyer}];
(\ Confirm_{Bank \rightarrow Seller} \parop 
Receipt_{Bank \rightarrow Buyer}\ )
\end{array}
\]
\end{example}


We are now ready to present the definition of our 
Orchestration Calculus extended with operators for dynamic updates:
\begin{displaymath}
  \begin{array} {rllllllllll}
    C ::= & ...\\
    \quad | \quad & X[C]^F            &\qquad (\text{scope}) & 
    \quad | \quad & \upd{X}{(r_1,\ldots,r_n)}{C_1,\ldots,C_n}      &\qquad (\text{update})
  \end{array}
\end{displaymath} 
where $F$ is either $A$, denoting an active (running) scope, or $\varepsilon$, denoting a scope still to be started ($\varepsilon$ is omitted in the following).
%
%
$X[C]^F$ denotes an endpoint scope named $X$
executing $C$. $F$ is a flag distinguishing scopes whose
execution has already begun ($A$) from scopes which have
not started yet ($\varepsilon$). In order to become active,
the endpoints involved in a scope, must synchronize.
Also when all participants in a scope completes their 
respective executions, a synchronisation is needed in order
to synchronously remove the scope. 
The update operator
$\upd{X}{(r_1,\ldots,r_n)}{C_1,\ldots,C_n}$ provides an update for scope named $X$, involving roles
$r_1,\ldots,r_n$. The new behaviour for role $r_i$ is $P_i$.

As in the previous sections, systems are of the form $[C]_r$, where $r$ is the name of
the role and $C$ its behaviour. Systems, denoted $P$, are 
obtained by 
composition of parallel
endpoints: 
\begin{displaymath}
  \begin{array} {rllllllllll}
    P\ ::= & [ C ]_r               &\qquad (\text{endpoint}) & 
    \quad | \quad & P \pps P     &\qquad (\text{parallel system})\\
  \end{array}
\end{displaymath} 

%

In this presentation, we do not formally define a semantics for
endpoints: we just point out that it should include labels
corresponding to all the labels of the semantics of choreographies, 
plus some additional labels corresponding to partial
activities, such as an input. We also highlight the fact that all
scopes which correspond to the same choreography scope evolve
together: their scope start transitions (transforming a scope from
inactive to active) are synchronized, as well as their scope end
transitions (removing it).  The fact that choreographies feature at
most one scope with a given name is instrumental in ensuring this
property.


We now discuss how to extend the notion of projection presented
in Definition~\ref{def:proj} for the case without updates.

\begin{definition}{\bf (Projection for choreographies with updates)}
The projection of a choreography $H$ on a role $r$, denoted by $\proj{H}{r}$, is defined 
as in Definition~\ref{def:proj} plus the clauses below for scopes and updates:
\begin{eqnarray*}
\proj{\upd{X}{r'}{H}}{r} & = & \begin{cases}\upd{X}{(r_1,\ldots,r_n)}{\proj{H}{r_1},\ldots,\proj{H}{r_n}}  
\text{with $\{r_1,\ldots,r_n\}= type(X)$} &
\!\!\!\!\text{if $r = r'$}\\
\one & 
\!\!\!\!\!\text{otherwise}\end{cases} \\
\proj{X:T[H]}{r} &= & \begin{cases} X[\proj{H}{r}] & \text{if $r \in type(X)$}\\
\one &  \text{otherwise} \end{cases}
\end{eqnarray*}
%
%
\end{definition}

\begin{example}
We now present the projection of the choreography 
in the Example~\ref{AdaptableBSB} (we omit unnecessary $\one$ terms):
\[
\begin{array}{l}
\ [\overline{Request}_{Seller} ; Offer ;
X[\overline{Payment}_{Bank}] ; Receipt
]_{Buyer}\ 
\pa \\
\ [Request ; (\overline{Offer}_{Buyer} \parop                \overline{PayDescr}_{Bank});Confirm]_{Seller} 
\ \pa \\
\ [PayDescr ; X[Payment]; (\overline{Receipt}_{Buyer} \parop  \overline{Confirm}_{Seller})]_{Bank}
\end{array}
\]
It is interesting to note that the projection clearly identifies where
a possible update of the payment should have an effect; namely,
only the roles $Buyer$ and $Bank$ are affected by the update
in precise parts of their behaviour.
For instance, if 
$\upd{X}{(Buyer,Bank)}{\big(\overline{VISAcode}_{Bank};VISAok\big),\big(VISAcode;\overline{VISAok}_{Buyer}\big)}$ 
is executed after the first $Request$
interaction occurs, then the system becomes:
\[
\begin{array}{l}
\ [Offer ;
X[\overline{VISAcode}_{Bank};VISAok] ; Receipt
]_{Buyer}\ 
\pa \\
\ [(\overline{Offer}_{Buyer} \parop                \overline{PayDescr}_{Bank});Confirm]_{Seller} 
\ \pa \\
\ [PayDescr ; X[VISAcode;\overline{VISAok}_{Buyer}]; (\overline{Receipt}_{Buyer} \parop  \overline{Confirm}_{Seller})]_{Bank}
\end{array}
\]
where the projections of the new protocol are precisely injected
in the behaviour of the affected roles.
\end{example}


Ideally, traces of the projected system
should correspond to the traces of the original
choreography. Actually, we conjecture that this occurs for
choreographies satisfying connectedness conditions 
obtained by extending those already discussed in Section~\ref{sec:connect}.
For instance it is necessary to consider also the new scope and update operators;
this can be done by adding to the 
auxiliary functions $\transI$ and $\transF$ rules to deal with the
novel constructs $X:T[C]$ and $\upd{X}{r}{C}$
(in the first case both functions should return the roles in $T$,
in the second case the unique role involved is $r$).
%
%
%

We finally point out two main aspects of the expected correspondence result
between choreographies and their projections in the case of the calculi
extended with dynamic updates.  First, labels
$\upd{X}{r}{H}$ 
of transitions of the choreography should be mapped to labels
of the transitions of the Orchestration Calculus
obtained by appropriate label projections.
Second, endpoint traces should not consider
unmatched input and output labels.

\section{Related Work and Conclusion}

Among our main contributions we can mention 
(i) the formalisation of the relationship between 
global choreographic descriptions and systems obtained 
as parallel compositions of peers,
(ii) the definition of suitable notions of contract refinement, and 
(iii) the proposal of mechanisms for dynamic 
updates for both the choreography and the orchestration calculi.
Concerning (i), we have defined
well-formedness for choreographies based on 
natural projection and the notion of implementation.
We have introduced the simple technique of obtaining orchestrations
by projection, defining it for communication actions and then 
exploiting omomorphism over all the process algebraic operators.
Moreover our approach leads to a more general notion of well-formedness
w.r.t. other approaches like, e.g.,~\cite{CHY07},
where it is defined in terms of three
connectedness constraints similar to those we
have defined in Section~\ref{sec:connect}.
Concerning (ii), our main contribution is related to 
the idea of refining all peers guaranteeing that
all of them continue to reach local success. 
This differs from other popular approaches,
like those initiated by Fournet et al.~\cite{FHRR04} 
or Padovani et al.~\cite{CCLP06}, where the focus is 
on the success of one specific peer (usually,
the so-called, {\em client}).
Concerning (iii), it is worth to mention
that~\cite{ivan} has been a source of inspiration
for the present work:
the main difference is 
our choice of expressing adaptation in terms of scopes 
and code update constructs, rather than using rules.
This approach appears more adequate for the definition 
of a general theory of behavioural typing to be used 
on more general languages where 
multiple protocols/choreographies 
can interleave inside the same program. 

Now some remarks concerning future work.
We are working on applying the theory of updatable choreographies/orchestrations in the context of session types for typing a concrete language with session spawning, where choreographies play the role of global types attached to sessions and we use orchestrations for checking, via typing rules, that the code actually conforms with the specified global types.
In this context, extending  our contract refinement theory to updatable choreographies/orchestrations (thus getting updatable behavioural contracts) would make it possible to define a notion of semantic subtyping.
We also plan to work on 
the complete characterization of compliance testing: work in this direction has been done in \cite{BH13} where however testing is characterized only for controllable processes (i.e. processes for which there exists a compliant test) and fairness is not considered (a testing notion similar to our compliance testing is considered,  where both the test and the system under test must succeed, but in the flavour of traditional testing of \cite{DH84} without assuming fairness).

\newcommand{\etalchar}[1]{$^{#1}$}

\providecommand{\urlalt}[2]{\href{#1}{#2}}
\providecommand{\doi}[1]{doi:\urlalt{http://dx.doi.org/#1}{#1}}


\end{document}